\documentstyle[12pt,dina4]{article}
\def\title #1 {
   \headsep=1.0in
   \baselineskip=30pt
			\begin{center}
   {\titlebold #1}
   \end{center}
			\vskip .75in }
\def\author #1 {
   \baselineskip=30pt
   \begin{center}
   {\timeslarge #1}
   \end{center}
			\vskip .25in }
\def\address #1 {
   \baselineskip=24pt
   \begin{center}
   {\timesitalic #1}
   \end{center}
   \vskip 1.0in }

\def\disp {\displaystyle}

\def\conj #1 {\overline #1}

\def\be {\begin{equation}}
\def\ee {\end{equation}}

\def\ba {\begin{array}}
\def\ea {\end{array}}
\def\bea {\begin{eqnarray}}
\def\eea {\end{eqnarray}}

\def\et {$$}

\def\etn {$$}

\def\ett {$$}

\def\ettn{$$}

\def\eqn#1 {(\ref{#1}) }

\newdimen\twoeqncolwidth
\newdimen\twoeqncolwidtha
\newdimen\twoeqncolwidthb
\newdimen\twoeqncolsep
\newdimen\twoeqnlinset
\def\twoeqn#1&#2\et{
   \hbox to\twoeqnlinset{\hfil}
   \hbox to\twoeqncolwidth{$\disp#1$\hfil}
   \hbox to\twoeqncolsep{\hfil}
   \hbox to\twoeqncolwidth{$\disp#2$\hfil}\eqno{\rm (\theequation)}$$}
\def\twoeqnt#1&#2\ett{
   \hbox to\twoeqnlinset{\hfil}
   \hbox to\twoeqncolwidtha{$\disp#1$\hfil}
   \hbox to\twoeqncolsep{\hfil}
   \hbox to\twoeqncolwidthb{$\disp#2$\hfil}\eqno{\rm (\theequation)}$$}
\def\twoeqnn#1&#2\etn{
   \hbox to\twoeqnlinset{\hfil}
   \hbox to\twoeqncolwidth{$\disp#1$\hfil}
   \hbox to\twoeqncolsep{\hfil}
   \hbox to\twoeqncolwidth{$\disp#2$\hfil}\eqno\phantom{\rm
(\theequation)}$$}
\def\twoeqntn#1&#2\ettn{
   \hbox to\twoeqnlinset{\hfil}
   \hbox to\twoeqncolwidtha{$\disp#1$\hfil}
   \hbox to\twoeqncolsep{\hfil}
   \hbox to\twoeqncolwidthb{$\disp#2$\hfil}\eqno\phantom{\rm
(\theequation)}$$}

\def\rawpicture #1 by #2 (#3){
  \vbox to #2{
    \hrule width #1 height 0pt depth 0pt
    \vfill
    \special{picture #3}     }
  }
\def\scaledpicture #1 by #2 (#3 scaled #4){{
  \dimen0=#1 \dimen1=#2
  \divide\dimen0 by 1000 \multiply\dimen0 by #4
  \divide\dimen1 by 1000 \multiply\dimen1 by #4
  \rawpicture \dimen0 by \dimen1 (#3 scaled #4)}
  }

\def\beginparmode{\endmode
  \begingroup \def\endmode{\par\endgroup}}
\let\endmode=\par
{\obeylines\gdef\
{}}

\def\body			  {\beginparmode}						
\def\head#1{			  \goodbreak\vskip 0.5truein	  {\immediate\write16{#1}
      \uppercase{#1}\par}
   \nobreak\vskip 0.25truein\nobreak}

\def\itemj{\par\hang\textindent}
\def\beginitems{
\par\medskip\bgroup\def\i##1 {\itemj{##1}}\def\ii##1 {\itemitem{##1}}
\leftskip=36pt\parskip=0pt}
\def\enditems{\par\egroup}

\def\beneathrel#1\under#2{\mathrel{\mathop{#2}\limits_{#1}}}

\def\refto#1{[#1]}		
\def\references			  {\head{REFERENCES}		   \beginparmode
   \frenchspacing \parindent=0pt \leftskip=1truecm
   \parskip=8pt plus 3pt \everypar{\hangindent=\parindent}}

\gdef\refis#1{\itemj{#1.\ }}			
\gdef\journal#1, #2, #3, 1#4#5#6{		    {\sl #1~}{\bf #2} (1#4#5#6), #3 }

\def\annp{\journal Ann. Phys. (N.Y.), }

\def\Annp{\journal Ann. Physik, }

\def\aspm{\journal Advanced Studies in Pure Mathematics, }

\def\cmp{\journal Comm. Math. Phys., }

\def\eurolett{\journal Europhysics Lett., }

\def\ijmpa{\journal Int. J. Mod. Phys. A, }

\def\ijmpb{\journal Int. J. Mod. Phys. B, }

\def\jappp{\journal J. Appl. Phys., }

\def\jphc{\journal J. Physique C, }

\def\jpI{\journal J. Physique I, }

\def\jpcoll{\journal J. Physique Coll, }

\def\jcr{\journal J. Chem. Res., }

\def\jetp{\journal Sov. Phys. JETP, }

\def\jetpl{\journal JETP Lett., }

\def\jpj{\journal J. Phys. Soc. Japan, }

\def\jmp{\journal J. Math. Phys., }

\def\jpa{\journal J. Phys. A, }

\def\jpc{\journal J. Phys. C, }

\def\jpcon{\journal J. Phys.: Condens. Matter, }

\def\ptp{\journal Prog. Theor. Phys., }

\def\jetp{\journal Sov. Phys. JETP, }

\def\jsp{\journal J. Stat. Phys., }

\def\lmp{\journal Lett. Math. Phys., }

\def\lnp{\journal Lecture Notes in Physics, }

\def\mpla{\journal Mod. Phys. Lett. A, }

\def\npb{\journal Nucl. Phys. B, }

\def\physica{\journal Physica, }

\def\pla{\journal Phys. Lett. A, }

\def\plb{\journal Phys. Lett. B, }

\def\prep{\journal Physics Reports, }

\def\pra{\journal Phys. Rev. A, }

\def\prb{\journal Phys. Rev. B, }

\def\prl{\journal Phys. Rev. Lett., }

\def\prs{\journal Proc. Roy. Soc. (London) A, }

\def\pr{\journal Phys. Rev., }

\def\rmp{\journal Rev. Mod. Phys., }

\def\sjnp{\journal Sov. J. Nucl. Phys., }

\def\tmp{\journal Theor. Math. Phys., }

\def\zpb{\journal Z. Phys. B, }

\def\zp{\journal Z. Phys., }

\def\reff#1{Ref.~#1}			\def\Reff#1{Ref.~#1}			\def\[#1]{[\refcite{#1}]}
\def\refcite#1{{#1}}
														\def\(#1){(\call{#1})}
\def\call#1{{#1}}
\def\taghead#1{}
\def\frac#1#2{{#1 \over #2}}

\def\sla{\raise.15ex\hbox{$/$}\kern-.57em}
\def\leaderfill{\leaders\hbox to 1em{\hss.\hss}\hfill}
\def\twiddle{\lower.9ex\rlap{$\kern-.1em\scriptstyle\sim$}}
\def\bigtwiddle{\lower1.ex\rlap{$\sim$}}
\def\gtwid{\mathrel{\raise.3ex\hbox{$>$\kern-.75em\lower1ex\hbox{$\sim$}}}}
\def\ltwid{\mathrel{\raise.3ex\hbox{$<$\kern-.75em\lower1ex\hbox{$\sim$}}}}
\def\square{\kern1pt\vbox{\hrule height 1.2pt\hbox{\vrule width 1.2pt\hskip
3pt
   \vbox{\vskip 6pt}\hskip 3pt\vrule width 0.6pt}\hrule height
0.6pt}\kern1pt}
\def\tdot#1{\mathord{\mathop{#1}\limits^{\kern2pt\ldots}}}

\def\pmb#1{\setbox0=\hbox{#1}  \kern-.025em\copy0\kern-\wd0
  \kern  .05em\copy0\kern-\wd0
  \kern-.025em\raise.0433em\box0 }

\def\refto#1{[#1]}		
\def\references			  {\section*{References}		   \beginparmode
   \frenchspacing \parindent=0pt \leftskip=1truecm
   \parskip=8pt plus 3pt \everypar{\hangindent=\parindent}}

\def\endreferences{\body}

\def\reff#1{Ref.~#1}			\def\Reff#1{Ref.~#1}			\def\[#1]{[\refcite{#1}]}
\def\refcite#1{{#1}}
\def\refeq#1{(\ref{#1})}

\catcode`@=11
\newcount\r@fcount \r@fcount=0
\newcount\r@fcurr
\immediate\newwrite\reffile
\newif\ifr@ffile\r@ffilefalse
\def\w@rnwrite#1{\ifr@ffile\immediate\write\reffile{#1}\fi\message{#1}}

\def\writer@f#1>>{}
\def\referencefile{  \r@ffiletrue\immediate\openout\reffile=\jobname.ref  \def\writer@f##1>>{\ifr@ffile\immediate\write\reffile    {\noexpand\refis{##1} = \csname r@fnum##1\endcsname =      \expandafter\expandafter\expandafter\strip@t\expandafter     \meaning\csname r@ftext\csname r@fnum##1\endcsname\endcsname}\fi}  \def\strip@t##1>>{}}

\def\citeall#1{\xdef#1##1{#1{\noexpand\refcite{##1}}}}
\def\refcite#1{\each@rg\citer@nge{#1}}	

\def\each@rg#1#2{{\let\thecsname=#1\expandafter\first@rg#2,\end,}}
\def\first@rg#1,{\thecsname{#1}\apply@rg}	\def\apply@rg#1,{\ifx\end#1\let\next=\relax\else,\thecsname{#1}\let\next=\apply@rg\fi\next}
\def\citer@nge#1{\citedor@nge#1-\end-}	
\def\citer@ngeat#1\end-{#1}
\def\citedor@nge#1-#2-{\ifx\end#2\r@featspace#1   \else\citel@@p{#1}{#2}\citer@ngeat\fi}	\def\citel@@p#1#2{\ifnum#1>#2{\errmessage{Reference range #1-#2\space is
bad.}
    \errhelp{If you cite a series of references by the notation M-N, then M
and
    N must be integers, and N must be greater than or equal to M.}}\else {\count0=#1\count1=#2\advance\count1
by1\relax\expandafter\r@fcite\the\count0,  \loop\advance\count0 by1\relax    \ifnum\count0<\count1,\expandafter\r@fcite\the\count0,  \repeat}\fi}

\def\r@featspace#1#2 {\r@fcite#1#2,}	\def\r@fcite#1,{\ifuncit@d{#1}    \newr@f{#1}    \expandafter\gdef\csname r@ftext\number\r@fcount\endcsname                     {\message{Reference #1 to be supplied.}                      \writer@f#1>>#1 to be supplied.\par} \fi \csname r@fnum#1\endcsname}
\def\ifuncit@d#1{\expandafter\ifx\csname r@fnum#1\endcsname\relax}\def\newr@f#1{\global\advance\r@fcount by1    \expandafter\xdef\csname r@fnum#1\endcsname{\number\r@fcount}}

\let\r@fis=\refis			\def\refis#1#2#3\par{\ifuncit@d{#1}   \newr@f{#1}   \w@rnwrite{Reference #1=\number\r@fcount\space is not cited up to
now.}\fi  \expandafter\gdef\csname r@ftext\csname r@fnum#1\endcsname\endcsname  {\writer@f#1>>#2#3\par}}

\def\ignoreuncited{   \def\refis##1##2##3\par{\ifuncit@d{##1}     \else\expandafter\gdef\csname r@ftext\csname
r@fnum##1\endcsname\endcsname     {\writer@f##1>>##2##3\par}\fi}}

\def\r@ferr{\endreferences\errmessage{I was expecting to see
\noexpand\endreferences before now;  I have inserted it here.}}
\let\r@ferences=\references
\def\references{\r@ferences\def\endmode{\r@ferr\par\endgroup}}

\let\endr@ferences=\endreferences
\def\endreferences{\r@fcurr=0  {\loop\ifnum\r@fcurr<\r@fcount    \advance\r@fcurr by
1\relax\expandafter\r@fis\expandafter{\number\r@fcurr}    \csname r@ftext\number\r@fcurr\endcsname  \repeat}\gdef\r@ferr{}\endr@ferences}

\let\r@fend=\endpaper\gdef\endpaper{\ifr@ffile
\immediate\write16{Cross References written on []\jobname.REF.}\fi\r@fend}

\catcode`@=12

\def\reftorange#1#2#3{[\refcite{#1}--\setbox0=\hbox{\refcite{#2}}\refcite{#3}]}

\citeall\refto		\citeall\reff		\citeall\Reff		

\ignoreuncited

\renewcommand{\title}[1]{\large\bf \mbox{}\\ \mbox{}\\ \mbox{}\\ \mbox{}\\
     #1\bigskip\medskip\\} 
\renewcommand{\author}[1]{\large #1\\ \smallskip}
\renewcommand{\address}[1]{{\narrower\normalsize\it #1\\}\bigskip}
\renewenvironment{abstract}{\narrower\small}{\par\normalsize\bigskip}

\catcode`\@=11

\font\twelvemsx=msxm10 scaled \magstep1
\font\tenmsx=msxm10
\font\sevenmsx=msxm7

\font\twelvemsy=msym10 scaled \magstep1
\font\tenmsy=msym10
\font\sevenmsy=msym7

\newfam\msxfam
\newfam\msyfam
\textfont\msxfam=\twelvemsx  
\scriptfont\msxfam=\tenmsx
  \scriptscriptfont\msxfam=\sevenmsx
\textfont\msyfam=\twelvemsy  
\scriptfont\msyfam=\tenmsy
  \scriptscriptfont\msyfam=\sevenmsy

\def\hexnumber@#1{\ifcase#1 0\or1\or2\or3\or4\or5\or6\or7\or8\or9\or
	A\or B\or C\or D\or E\or F\fi }

\font\twelveeuf=eufm10 scaled \magstep1
\font\teneuf=eufm10
\font\seveneuf=eufm7

\newfam\euffam
\textfont\euffam=\twelveeuf
\scriptfont\euffam=\teneuf
\scriptscriptfont\euffam=\seveneuf
\def\frak{\relaxnext@\ifmmode\let\next\frak@\else
 \def\next{\Err@{Use \string\frak\space only in math mode}}\fi\next}
\def\goth{\relaxnext@\ifmmode\let\next\frak@\else
 \def\next{\Err@{Use \string\goth\space only in math mode}}\fi\next}
\def\frak@#1{{\frak@@{#1}}}
\def\frak@@#1{\noaccents@\fam\euffam#1}

\edef\msx@{\hexnumber@\msxfam}
\edef\msy@{\hexnumber@\msyfam}

\mathchardef\boxdot="2\msx@00
\mathchardef\boxplus="2\msx@01
\mathchardef\boxtimes="2\msx@02
\mathchardef\square="0\msx@03
\mathchardef\blacksquare="0\msx@04
\mathchardef\centerdot="2\msx@05
\mathchardef\lozenge="0\msx@06
\mathchardef\blacklozenge="0\msx@07
\mathchardef\circlearrowright="3\msx@08
\mathchardef\circlearrowleft="3\msx@09
\mathchardef\rightleftharpoons="3\msx@0A
\mathchardef\leftrightharpoons="3\msx@0B
\mathchardef\boxminus="2\msx@0C
\mathchardef\Vdash="3\msx@0D
\mathchardef\Vvdash="3\msx@0E
\mathchardef\vDash="3\msx@0F
\mathchardef\twoheadrightarrow="3\msx@10
\mathchardef\twoheadleftarrow="3\msx@11
\mathchardef\leftleftarrows="3\msx@12
\mathchardef\rightrightarrows="3\msx@13
\mathchardef\upuparrows="3\msx@14
\mathchardef\downdownarrows="3\msx@15
\mathchardef\upharpoonright="3\msx@16

\mathchardef\downharpoonright="3\msx@17
\mathchardef\upharpoonleft="3\msx@18
\mathchardef\downharpoonleft="3\msx@19
\mathchardef\rightarrowtail="3\msx@1A
\mathchardef\leftarrowtail="3\msx@1B
\mathchardef\leftrightarrows="3\msx@1C
\mathchardef\rightleftarrows="3\msx@1D
\mathchardef\Lsh="3\msx@1E
\mathchardef\Rsh="3\msx@1F
\mathchardef\rightsquigarrow="3\msx@20
\mathchardef\leftrightsquigarrow="3\msx@21
\mathchardef\looparrowleft="3\msx@22
\mathchardef\looparrowright="3\msx@23
\mathchardef\circeq="3\msx@24
\mathchardef\succsim="3\msx@25
\mathchardef\gtrsim="3\msx@26
\mathchardef\gtrapprox="3\msx@27
\mathchardef\multimap="3\msx@28
\mathchardef\therefore="3\msx@29
\mathchardef\because="3\msx@2A
\mathchardef\doteqdot="3\msx@2B

\mathchardef\triangleq="3\msx@2C
\mathchardef\precsim="3\msx@2D
\mathchardef\lesssim="3\msx@2E
\mathchardef\lessapprox="3\msx@2F
\mathchardef\eqslantless="3\msx@30
\mathchardef\eqslantgtr="3\msx@31
\mathchardef\curlyeqprec="3\msx@32
\mathchardef\curlyeqsucc="3\msx@33
\mathchardef\preccurlyeq="3\msx@34
\mathchardef\leqq="3\msx@35
\mathchardef\leqslant="3\msx@36
\mathchardef\lessgtr="3\msx@37
\mathchardef\backprime="0\msx@38
\mathchardef\risingdotseq="3\msx@3A
\mathchardef\fallingdotseq="3\msx@3B
\mathchardef\succcurlyeq="3\msx@3C
\mathchardef\geqq="3\msx@3D
\mathchardef\geqslant="3\msx@3E
\mathchardef\gtrless="3\msx@3F
\mathchardef\sqsubset="3\msx@40
\mathchardef\sqsupset="3\msx@41
\mathchardef\vartriangleright="3\msx@42
\mathchardef\vartriangleleft="3\msx@43
\mathchardef\trianglerighteq="3\msx@44
\mathchardef\trianglelefteq="3\msx@45
\mathchardef\bigstar="0\msx@46
\mathchardef\between="3\msx@47
\mathchardef\blacktriangledown="0\msx@48
\mathchardef\blacktriangleright="3\msx@49
\mathchardef\blacktriangleleft="3\msx@4A
\mathchardef\vartriangle="0\msx@4D
\mathchardef\blacktriangle="0\msx@4E
\mathchardef\triangledown="0\msx@4F
\mathchardef\eqcirc="3\msx@50
\mathchardef\lesseqgtr="3\msx@51
\mathchardef\gtreqless="3\msx@52
\mathchardef\lesseqqgtr="3\msx@53
\mathchardef\gtreqqless="3\msx@54
\mathchardef\Rrightarrow="3\msx@56
\mathchardef\Lleftarrow="3\msx@57
\mathchardef\veebar="2\msx@59
\mathchardef\barwedge="2\msx@5A
\mathchardef\doublebarwedge="2\msx@5B
\mathchardef\angle="0\msx@5C
\mathchardef\measuredangle="0\msx@5D
\mathchardef\sphericalangle="0\msx@5E
\mathchardef\varpropto="3\msx@5F
\mathchardef\smallsmile="3\msx@60
\mathchardef\smallfrown="3\msx@61
\mathchardef\Subset="3\msx@62
\mathchardef\Supset="3\msx@63
\mathchardef\Cup="2\msx@64

\mathchardef\Cap="2\msx@65

\mathchardef\curlywedge="2\msx@66
\mathchardef\curlyvee="2\msx@67
\mathchardef\leftthreetimes="2\msx@68
\mathchardef\rightthreetimes="2\msx@69
\mathchardef\subseteqq="3\msx@6A
\mathchardef\supseteqq="3\msx@6B
\mathchardef\bumpeq="3\msx@6C
\mathchardef\Bumpeq="3\msx@6D
\mathchardef\lll="3\msx@6E

\mathchardef\ggg="3\msx@6F

\mathchardef\circledS="0\msx@73
\mathchardef\pitchfork="3\msx@74
\mathchardef\dotplus="2\msx@75
\mathchardef\backsim="3\msx@76
\mathchardef\backsimeq="3\msx@77
\mathchardef\complement="0\msx@7B
\mathchardef\intercal="2\msx@7C
\mathchardef\circledcirc="2\msx@7D
\mathchardef\circledast="2\msx@7E
\mathchardef\circleddash="2\msx@7F
\def\ulcorner{\delimiter"4\msx@70\msx@70 }
\def\urcorner{\delimiter"5\msx@71\msx@71 }
\def\llcorner{\delimiter"4\msx@78\msx@78 }
\def\lrcorner{\delimiter"5\msx@79\msx@79 }
\def\yen{\mathhexbox\msx@55 }
\def\checkmark{\mathhexbox\msx@58 }
\def\circledR{\mathhexbox\msx@72 }
\def\maltese{\mathhexbox\msx@7A }
\mathchardef\lvertneqq="3\msy@00
\mathchardef\gvertneqq="3\msy@01
\mathchardef\nleq="3\msy@02
\mathchardef\ngeq="3\msy@03
\mathchardef\nless="3\msy@04
\mathchardef\ngtr="3\msy@05
\mathchardef\nprec="3\msy@06
\mathchardef\nsucc="3\msy@07
\mathchardef\lneqq="3\msy@08
\mathchardef\gneqq="3\msy@09
\mathchardef\nleqslant="3\msy@0A
\mathchardef\ngeqslant="3\msy@0B
\mathchardef\lneq="3\msy@0C
\mathchardef\gneq="3\msy@0D
\mathchardef\npreceq="3\msy@0E
\mathchardef\nsucceq="3\msy@0F
\mathchardef\precnsim="3\msy@10
\mathchardef\succnsim="3\msy@11
\mathchardef\lnsim="3\msy@12
\mathchardef\gnsim="3\msy@13
\mathchardef\nleqq="3\msy@14
\mathchardef\ngeqq="3\msy@15
\mathchardef\precneqq="3\msy@16
\mathchardef\succneqq="3\msy@17
\mathchardef\precnapprox="3\msy@18
\mathchardef\succnapprox="3\msy@19
\mathchardef\lnapprox="3\msy@1A
\mathchardef\gnapprox="3\msy@1B
\mathchardef\nsim="3\msy@1C
\mathchardef\ncong="3\msy@1D

\mathchardef\varsubsetneq="3\msy@20
\mathchardef\varsupsetneq="3\msy@21
\mathchardef\nsubseteqq="3\msy@22
\mathchardef\nsupseteqq="3\msy@23
\mathchardef\subsetneqq="3\msy@24
\mathchardef\supsetneqq="3\msy@25
\mathchardef\varsubsetneqq="3\msy@26
\mathchardef\varsupsetneqq="3\msy@27
\mathchardef\subsetneq="3\msy@28
\mathchardef\supsetneq="3\msy@29
\mathchardef\nsubseteq="3\msy@2A
\mathchardef\nsupseteq="3\msy@2B
\mathchardef\nparallel="3\msy@2C
\mathchardef\nmid="3\msy@2D
\mathchardef\nshortmid="3\msy@2E
\mathchardef\nshortparallel="3\msy@2F
\mathchardef\nvdash="3\msy@30
\mathchardef\nVdash="3\msy@31
\mathchardef\nvDash="3\msy@32
\mathchardef\nVDash="3\msy@33
\mathchardef\ntrianglerighteq="3\msy@34
\mathchardef\ntrianglelefteq="3\msy@35
\mathchardef\ntriangleleft="3\msy@36
\mathchardef\ntriangleright="3\msy@37
\mathchardef\nleftarrow="3\msy@38
\mathchardef\nrightarrow="3\msy@39
\mathchardef\nLeftarrow="3\msy@3A
\mathchardef\nRightarrow="3\msy@3B
\mathchardef\nLeftrightarrow="3\msy@3C
\mathchardef\nleftrightarrow="3\msy@3D
\mathchardef\divideontimes="2\msy@3E
\mathchardef\varnothing="0\msy@3F
\mathchardef\nexists="0\msy@40
\mathchardef\mho="0\msy@66
\mathchardef\eth="0\msy@67
\mathchardef\eqsim="3\msy@68
\mathchardef\beth="0\msy@69
\mathchardef\gimel="0\msy@6A
\mathchardef\daleth="0\msy@6B
\mathchardef\lessdot="3\msy@6C
\mathchardef\gtrdot="3\msy@6D
\mathchardef\ltimes="2\msy@6E
\mathchardef\rtimes="2\msy@6F
\mathchardef\shortmid="3\msy@70
\mathchardef\shortparallel="3\msy@71
\mathchardef\smallsetminus="2\msy@72
\mathchardef\thicksim="3\msy@73
\mathchardef\thickapprox="3\msy@74
\mathchardef\approxeq="3\msy@75
\mathchardef\succapprox="3\msy@76
\mathchardef\precapprox="3\msy@77
\mathchardef\curvearrowleft="3\msy@78
\mathchardef\curvearrowright="3\msy@79
\mathchardef\digamma="0\msy@7A
\mathchardef\varkappa="0\msy@7B
\mathchardef\hslash="0\msy@7D
\mathchardef\hbar="0\msy@7E
\mathchardef\backepsilon="3\msy@7F
\def\Bbb@@#1{\fam\msyfam#1}
\def\Bbb@#1{{\Bbb@@{#1}}}
\def\Bbb{\Bbb@}

\catcode`\@=12

\font\twelvemsx=msxm10 scaled \magstep1
\font\twelvemsy=msym10 scaled \magstep1
\font\twelveeuf=eufm10 scaled \magstep1
\textfont\msxfam=\twelvemsx
\textfont\msyfam=\twelvemsy
\textfont\euffam=\twelveeuf
\def\frak#1{\mbox{\twelveeuf #1}}

\def\phi{\varphi}
\def\-{{\bf --}}

\newcommand{\la}{\lambda}
\newcommand{\eps}{\varepsilon}

\newcommand{\cc}{{\hbox to 8pt {\hfill\vrule height 6.5pt \kern-2.6pt {\rm C}
                  \hfill}}}

\newcounter{num}

\def\cre#1#2{c^+_{#1 #2}}
\def\ann#1#2{c_{#1 #2}}
\def\num#1#2{n_{#1 #2}}

\def\e{{\rm e}}

\def\qLam#1#2{q^{#1}\left({i\over 2}#2 \eta\right)}
\def\qLam2#1#2{q^{#1}\left({i\over 2}#2 {\eta\over 2}\right)}

\begin{document}
\begin{center}

\titlepage

\title{Critical exponents of a multicomponent anisotropic
\mbox{\large\bf $t-J$ model in one dimension}\footnote{
Work performed within the research program of the 
Sonderforschungsbereich 341, K\"oln-Aachen-J\"ulich}}

\vskip1cm

\author{R. Z. Bariev\footnote{Permanent address: The Kazan Physico-Technical
Institute of the Russian Academy of Sciences, Kazan 420029, Russia}, 
A. Kl\"umper, A. Schadschneider\footnote{Present address: 
Institute for Theoretical Physics, 
State University of New York at
Stony Brook, Stony Brook, N Y 11794-3840, U.S.A.}, J. Zittartz}
\address{Institut f\"ur
Theoretische Physik,
Universit\"at zu K\"oln, Z\"ulpicher Str. 77,\\ 
D-50937 K\"oln, Germany.
\footnote{Email: kluemper@thp.uni-koeln.de, as@thp.uni-koeln.de,
zitt@thp.uni-koeln.de}}
\end{center}

\vskip1cm

\begin{abstract}
A recently presented anisotropic generalization of the multicomponent
supersymmetric $t-J$ model in one dimension is investigated. This
model of fermions with general spin-$S$ is solved by Bethe ansatz for
the ground state and the low-lying excitations.  Due to the anisotropy
of the interaction the model possesses $2S$ massive modes and one
single gapless excitation.  The physical properties indicate the
existence of Cooper-type multiplets of $2S+1$ fermions with finite
binding energy.  The critical behaviour is described by a $c=1$
conformal field theory with continuously varying exponents depending
on the particle density.  There are two distinct regimes of the phase
diagram with dominating density-density and multiplet-multiplet
correlations, respectively. The effective mass of the charge carriers is
calculated. In comparison to the limit of isotropic interactions the mass
is strongly enhanced in general.

\end{abstract}

\vspace{1cm}

\today

\bibliographystyle{alpha}

\newpage

\section{Introduction}
The $t-J$ model is one of the most fundamental models of strongly
correlated electron systems. It has been studied intensively in recent 
years since its connection with high $T_c$ superconductivity was pointed out
\refto{And87,ZhangR88}. 
It has been conjectured \refto{And90}
that one and two-dimensional variants of this model 
have properties in common. Therefore it is important to have exact
results available in one dimension for integrable coupling parameters.

The $t-J$ model describes electrons with nearest-neighbour hopping and
spin-exchange interaction with the constraint that no site is allowed to be 
occupied by more than one particle. The one-dimensional model for spin-1/2
particles was found to be integrable at the supersymmetric point 
\reftorange{Suth75}{Schlott87}{BaresB90} and the 
critical exponents for the long-distance asymptotics of its correlation 
functions were calculated \refto{KawY90}. 
These results were extended to a $t-J$ model of
fermions of arbitrary spin $S$ with supersymmetric interaction
\reftorange{LeeS88}{Schlott92}{Kaw93}. Recently,
the integrability of an anisotropic generalization of the multicomponent
supersymmetric $t-J$ model has been shown
\refto{Bariev94b,FoersK93}. The Hamiltonian reads
\bea
H=&-&\sum_{j,s}{\cal P}\left(\cre{j}{s}\ann{j+1}{s}+\cre{j+1}{s}\ann{j}{s}
\right){\cal P}\cr
&+&\sum_{j,s,s'}\left\{\cre{j}{s}\ann{j}{s'}\cre{j+1}{s'}\ann{j+1}{s}
-\exp[\hbox{sign}(s'-s)\eta]\num{j}{s}\num{j+1}{s'}\right\},\label{hamil}
\eea
where $\ann{j}{s}$ annihilates a fermion at site $j$ with spin component $s$. 
We assume $s=$ 1, 2, ..., $2S+1$ for convenience. $\num{j}{s}$ denotes the
number operator and ${\cal P}$ the projection operator onto the subspace of 
at most singly occupied sites. For $\eta=0$ the interaction is
supersymmetric. In the general case the Hamiltonian is spl$_q(2S+1,1)$
invariant and $\eta>0$ is a measure of the anisotropy.

In this paper we calculate the critical exponents of the correlation 
functions for the integrable anisotropic $t-J$ model of fermions with
arbitrary spin $S$ \refeq{hamil}. For $S=1/2$ some of these results were 
presented in \refto{Bariev94a}. The paper is organized as follows. In 
section 2
the Bethe ansatz solution is presented and the ground state and excitations
are discussed. In section 3 the density-density and superconducting 
correlation functions are investigated, and the transport masses are 
calculated. The paper ends with a discussion of the results.

\section{The Bethe Ansatz}

The general Bethe ansatz solution of the eigenvalue problem for Hamiltonian
\refeq{hamil} has been obtained 
on the basis of the Perk-Schultz model \reftorange{PerkS81}{BDV82}{Schul83} 
in \refto{Bariev94b}
from which we quote the relevant equations. The eigenstates are 
characterized by sets of rapidities $\la_j^{(0)}$ ($j=$ 1, 2, ..., $N$) for 
the electrons which total number is $N$, and by spin
rapidities $\la_\alpha^{(l)}$ ($l=$ 1, 2, ..., $2S$; $\alpha=$ 1, 2, ...,
$M_l$). The eigenvalue problem
reduces to the task of solving the following Bethe ansatz equations
\bea
\left[{\sin\left(\la_j^{(0)}-i\eta/2\right)\over
\sin\left(\la_j^{(0)}+i\eta/2\right)}\right]^L&=&
\prod_{\alpha=1}^{M_1}
{\sin\left(\la_j^{(0)}-\la_\alpha^{(1)}-i\eta/2\right)\over
\sin\left(\la_j^{(0)}-\la_\alpha^{(1)}+i\eta/2\right)},\cr
\prod_{\beta=1}^{M_l}
{\sin\left(\la_\alpha^{(l)}-\la_\beta^{(l)}-i\eta\right)\over
\sin\left(\la_\alpha^{(l)}-\la_\beta^{(l)}+i\eta\right)}
&=&-\prod_{\beta=1}^{M_{l-1}}
{\sin\left(\la_\alpha^{(l)}-\la_\beta^{(l-1)}-i\eta/2\right)\over
\sin\left(\la_\alpha^{(l)}-\la_\beta^{(l-1)}+i\eta/2\right)}
\cdot\cr&&\phantom{-}
\prod_{\beta=1}^{M_{l+1}}
{\sin\left(\la_\alpha^{(l)}-\la_\beta^{(l+1)}-i\eta/2\right)\over
\sin\left(\la_\alpha^{(l)}-\la_\beta^{(l+1)}+i\eta/2\right)},
\label{BA}
\eea
where $L$ is the number of lattice sites and we have set $M_0=N$ and
$M_{2S+1}=0$. (The number of particles with spin component $m$ is conserved
and equal to $M_{2S+1-m}-M_{2S+2-m}$.) 
The total energy and momentum of the system are given in terms of the 
electron rapidities $\la_j^{(0)}$
\bea
E&=&-2 N \cosh\eta+2\sinh^2\eta\sum_{j=1}^N{1\over\cosh\eta-\cos 2\la_j^{(0)}},\cr
P&=&N\pi-\sum_{j=1}^N\Theta\left(2\la_j^{(0)};{\eta\over 2}\right),
\label{energ}
\eea
where we have used the definition
\be
\Theta(v;\eta):=-i\ln{\sin\left(i\eta-v/ 2\right)
\over\sin\left(i\eta+v/2\right)}=
2\arctan\left(\coth\eta \tan{v\over 2}\right).
\ee

We now discuss the patterns of distributions of rapidities satisfying 
\refeq{BA}. Due to the attractive nature of the interaction the ground
state is decribed by bound complexes of rapidities $\la^{(l)}$ ($l=$ 0, 1, ...,
$2S-1$) which can be parametrized conveniently by the (real) rapidities
$\la^{(2S)}$. (In this paragraph subscripts are dropped). 
In order to accomodate for the groundstate we assume that
the number of electrons $N$ is a multiple of $2S+1$, $N=M(2S+1)$. Then we
have $M_{2S}=M$ real rapidities $\la^{(2S)}$ and for each such $\la^{(2S)}$
there are $2S+1-l$ complex spin rapidities $\la^{(l)}$ given by
\be
\la^{(2S)}+i{p\over 2}\eta,\quad p=-(2S-l),\,-(2S-l-2),\,...,\,(2S-l-2),\,
(2S-l).\label{string}
\ee
In particular we have $M_l=M(2S+1-l)$ from which we conclude that 
the particles 
are equally distributed over all spin components, i.e. the state is a singlet.

Using \refeq{string} the equations \refeq{BA} are straightforwardly reduced 
to one set of $M$ equations for the parameters $v_\alpha=2\la_\alpha^{(2S)}$
\bea
L\,\Theta\left(v_\alpha;{2S+1\over 2}\eta\right)&=&2\pi I_\alpha+
\sum_{\beta=1}^M\theta(v_\alpha-v_\beta),\cr
\theta(v)&:=&\sum_{l=1}^{2S}\Theta(v;l\eta).
\label{SimpBA}
\eea
The numbers $I_\alpha$ are integer (half-integer) for even (odd) \ $2S(M-1)$.
In the groundstate these numbers are arranged symmetrically around zero.
The energy and momentum are given by
\be
E=\sum_{\alpha=1}^Me(v_\alpha),\qquad P=\sum_{\alpha=1}^Mp(v_\alpha),
\ee
with the following abbreviations
\bea
e(v)&:=&2\left\{{\sinh(2S+1)\eta \sinh\eta\over \cosh(2S+1)\eta-\cos v}
-(2S+1)\cosh\eta\right\}\cr
p(v)&:=&\pi-\Theta\left(v;{2S+1\over 2}\eta\right).
\eea

In the thermodynamic limit $L,\,M\to\infty$ with finite $M/L$ the parameters 
$v_\alpha$ fill an interval $[v_0,2\pi-v_0]$ uniformly with density $\sigma(v)$.
>From \refeq{SimpBA} we obtain an integral equation of Fredholm type for the 
distribution function $\sigma(v)$
\be
2\pi\sigma(v)=\Theta'\left(v;{2S+1\over 2}\eta\right)-\int_{v_0}^{2\pi-v_0}
\theta'(v-w)\sigma(w)dw,
\label{IntGl}
\ee
with $\Theta'(v,\eta)=\sinh 2\eta/(\cosh 2\eta-\cos v)$
and subsidiary condition
\be
\int_{v_0}^{2\pi-v_0}\sigma(v)dv={M\over L}={\rho\over 2S+1},
\ee
where $\rho$ is the particle density. The solution to these equations 
$\sigma(v)$ yields the ground state energy per site
\be
{E_0\over L}=\int_{v_0}^{2\pi-v_0}e(v) \sigma(v)dv.
\ee

A gapless mode of excitations is given by a redistribution of the
$v_\alpha$ parameters in \refeq{SimpBA} by keeping the string structure 
\refeq{string}. These excitations are of particle-hole type with
quasilinear dispersion. There are $2S$ massive modes of excitations
corresponding to the breaking of some complexes \refeq{string}. 
These energy states are not considered here in detail as they 
are irrelevant for the behaviour of the algebraically
decaying correlations. The integral equations defining these excitations
as well as analytic results for the gaps in
various limits of $\rho$ and $\eta$ can be found in the appendix.

\section{Critical exponents of the correlation functions and masses of
charge carriers}

In this section we present the results for the correlation functions 
which are
obtained from applications of conformal field theory to finite-size 
calculations performed as in \refto{BarievKSZ93}.
In the anisotropic
model we have only one gapless excitation in contrast to the isotropic
model with $2S+1$ such modes. As a result our model is described by a $c=1$
conformal field theory and the scaling dimensions of the primary fields are 
uniquely expressed in terms of the dressed charge
\reftorange{KadaB79}{Card86a,IKR89}{BogK89} as
\be
x=\left[{\Delta M\over 2\xi(v_0)}\right]^2+\left[\xi(v_0) d\right]^2,
\label{Dim}
\ee
where $\Delta M$ and $d$ label the quantum numbers which specify the massless 
excitations. $\Delta M$ is the change in the number of complexes 
\refeq{string} compared to the groundstate, $d$ is the number of complexes
excited from the left Fermi point to the right one. All other excitations
correspond to broken complexes \refeq{string} and have a gap. Therefore these
excitations do not affect the critical properties.

The dressed charge function $\xi(v)$ is given by the solution of the integral
equation associated with \refeq{IntGl}
\be
2\pi\xi(v)=2\pi-\int_{v_0}^{2\pi-v_0}\theta'(v-w)\xi(w)dw.
\ee

According to conformal field theory the long-distance asymptotics of the
equal-time correlation functions of primary scaling fields $\Phi(r)$ are
determined by the critical exponents $2x$
\be
\langle\Phi(r)\Phi(0)\rangle\simeq{\exp(-2id k_F r)\over r^{2x}},\quad
k_F={\rho\over 2S+1}\pi.
\ee
The dimensions of descendent fields differ from $x$ by integers $N^\pm$.

We first consider the density-density correlations. The leading contributions
to the algebraic decay of this correlator are given by formula \refeq{Dim}
for $d=0$, $(N^+,N^-)=(1,0)$ or $(0,1)$, and $d=1$, $N^\pm=0$, respectively.
This leads to the asymptotic form
\be
\langle\rho(r)\rho(0)\rangle\simeq\rho^2+A_1r^{-2}+A_2r^{-\alpha}\cos(2k_Fr),
\ee
where
\be
\rho(r)=\sum_{s=1}^{2S+1}\num{r}{s},\quad\alpha=2\left[\xi(v_0)\right]^2.
\ee

Turning to the superconducting correlation functions we discuss the 
correlations of complexes of 2S+1 particles
\be
C(r)=\prod_{s=1}^{2S+1}\ann{r}{s}.
\ee
In this case the operator changes the number of complexes by $\Delta M=1$ with
$d=0$ or $d=1/2$ for half-integer or integer spin $S$, respectively. We
thus obtain
\be
\langle C^+(r)C(0)\rangle\simeq B r^{-\beta},\quad \beta={1\over\alpha},
\quad\hbox{ for half-integer } S,
\ee
and
\be
\langle C^+(r)C(0)\rangle\simeq B r^{-\beta}\cos(k_F r),\quad 
\beta={1\over\alpha}+{\alpha\over 4}, \quad\hbox{ for integer } S.
\ee

In our one-dimensional system we have no superconductivity in the 
literal sense.
Our model does not have finite off-diagonal long-range order. However, the
superconducting correlations have a longer range than the density-density
correlations
provided $\beta<\alpha$. Analytically we find $\alpha(\rho=0)=2$ and 
$\alpha(\rho_{\rm max}=1)=2/(2S+1)^2$. This implies that for all $S$ there
is a density regime $[0,\rho_c]$ where the system has dominating superconducting
correlations, see Fig. 1. This property of the anisotropic model differs 
from the behaviour of the isotropic $t-J$ 
model where spin and charge rapidities are also 
classified as bound states, however, with arbitrarily small binding energy. 
Therefore, the latter system does not manifest superconducting properties.

The formation of complexes of $2S+1$ bound particles can be substantiated
through a study of the conduction properties, notably the transport masses of
the charge carriers. Following \refto{ShasS90,KawY91,BarievKSZ93} we determine
the charge stiffness (Drude weight) $D_c$ from the dressed charge $\xi$
\be
D_c={(2S+1)^2\over 2\pi}v_F \xi^2(v_0).
\ee
$v_F$ is the velocity of the gapless excitations 
\be
v_F={|\eps'(v_0)|\over 2\pi\sigma(v_0)},
\ee
and $\eps(v)$ is the dressed energy defined in \refeq{DressEnerg} in the 
appendix. Figs. 2 and 3 show numerical results for the charge stiffness 
and for the effective mass $m$ of the charge carriers per bare mass $m_e$ 
of the particles
\be
{m\over m_e}={D_c^0\over D_c},
\ee
where $D_c^0={2S+1\over\pi}\sin\pi{\rho\over 2S+1}$ 
is the charge stiffness of free fermions with spin $S$. 
In the strong-coupling limit
the charge stiffness is of order $\e^{-2S\eta}$
\be
D_c={(2S+1)^2\over\pi}\left(1-{2S\over 2S+1}\rho\right)
\sin{\pi\rho\over 2S+1-2S\rho}\e^{-2S\eta},
\ee
and therefore the transport masses are of order $\e^{2S\eta}$. The reason
for these large masses is the binding of the particles in complexes.
The hopping of particles from one 
lattice site to another is always accompanied by the breaking of the complex
to which it belongs, at the expense of the binding energy 
of order $\e^{2S\eta}$. 

In summary, we have obtained the spectral gaps, the transport masses, and
the exact critical exponents of the long-distance
asymptotics of the correlation functions of an anisotropic 
multicomponent $t-J$ model. In particular the algebraically
decaying density-density and superconducting correlations have been analyzed
and superconducting properties have been found. The one-particle Green's 
functions $\langle\cre{r}{s}\ann{0}{s}\rangle$ on the other hand
show exponential decay, because
the fermion operators $\ann{r}{s}$ change the total number of particles by
1 corresponding to an excitation with gap. As a consequence the distribution
function $\langle\num{k}{s}\rangle$ is non-singular. Therefore the model
under consideration is characterized as a (marginal) Luttinger liquid with 
a single mode of gapless charge excitations and $2S$ spin excitations with gap.

\renewcommand{\theequation}{A.\arabic{equation}}
\section*{Appendix}
\setcounter{equation}{0}

Here we give a brief account of the elementary excitations over the ground 
state which is characterized by a distribution of the $\la_j^{(0)}$ parameters
in strings of length $2S+1$. 

The first kind of excitations consists of a redistribution of these strings 
and is of particle-hole type. The relevant equation governing the distribution
of the string parameters $v_\alpha$ is \refeq{SimpBA} from which we derive
the integral equation for the dressed energy $\eps(v)$ in standard way 
\be
\eps(v)+{1\over 2\pi}\int_{v_0}^{2\pi-v_0}
\theta'(v-w)\eps(w)dw=e(v)+(2S+1)\mu,
\label{DressEnerg}
\ee
where $e(v)$ is the bare energy of $(2S+1)$-strings with rapidity $v$ and
we have introduced the chemical potential $\mu$ per particle. The chemical
potential is a function of the particle density $\rho$ ($=\rho(v_0)$) via
the condition
\be
\eps(v_0)=\eps(2\pi-v_0)=0.
\ee
The dressed energy $\eps(v)$ is negative for $v$ inside the interval 
$[v_0,2\pi-v_0]$ and positive for $v$ outside,
thus corresponding to excitations of hole- and particle-type, respectively.
The general particle-hole type excitations read
\bea
E-E_0&=&\sum_{v_p}\eps(v_p)-\sum_{v_n}\eps(v_n),\cr
P&=&\sum_{v_p}\pi(v_p)-\sum_{v_n}\pi(v_n),
\eea
where the momentum $\pi(v)$ is defined by
\be
\pi(v)=-{1\over 2\sinh\eta}\int_\pi^v\left[\eps(w)+(2S+1)(2\cosh\eta-\mu)
\xi(w)\right]dw.
\ee

There are $2S$ massive modes of excitations corresponding to the breaking
of some strings \refeq{string}. In the general situation the $\la_j^{(0)}$
are distributed in additional strings of length $m=$ 1, 2, ..., $2S$. Instead of 
\refeq{SimpBA} we obtain
\be
L\,\Theta\left(v_\alpha;{2S+1\over 2}\eta\right)=2\pi I_\alpha+
\sum_{\beta=1}^{M'}\theta(v_\alpha-v_\beta)
+\sum_{m=1}^{2S}\sum_\beta\theta_m(v_\alpha-v_\beta^{(m)}),
\label{SimpExc}
\ee
where the parameters $v_\beta^{(m)}$ are the rapidities of the strings of
length $m$ and we have used the definition
\be
\theta_m(v)=\sum_{l=(2S+1-m)/2}^{(2S-1+m)/2}\Theta(v;l\eta).
\ee
Applying standard transformations to \refeq{SimpExc} we arrive at the
energy-momentum dispersions of the massive modes in terms of the dressed 
energy $\eps(v)$
\bea
\eps_m(v)&=&2\left({\sinh m\eta\sinh\eta\over\cosh m\eta-\cos v}
-m\cosh\eta\right)+m\mu
-{1\over 2\pi}\int_{v_0}^{2\pi-v_0}\theta_m'(v-w)\eps(w)dw,\cr
p_m(v)&=&\pi-\Theta\left(v,{m\over 2}\eta\right)\,\cr
&&+{1\over 4\pi\sinh\eta}\int_{v_0}^{2\pi-v_0}\theta_m(v-w)
\left[\eps(w)+(2S+1)(2\cosh\eta-\mu)\xi(w)\right]dw.\cr
&&
\eea

The gap $\Delta_m=\epsilon_m(\pi)$ of the $m^{\rm th}$ excitation
is found to be $0$ and $[1-m/(2S+1)]\e^\eta$ in the limits of the
anisotropy parameter $\eta\to 0$
and $\eta\to\infty$, respectively. For density $\rho\to 0$ we find
\be
\Delta_m=2\sinh\eta\left({\sinh m\eta\over\cosh m\eta+1}
-{m\over 2S+1}{\sinh(2S+1)\eta\over\cosh(2S+1)\eta+1}\right),
\ee
and for maximum density $\rho\to 1$
\be
\Delta_m=2\sinh\eta\left({\sinh m\eta\over\cosh m\eta+1}
-{m\over 2S+1}-2\sum_{k=1}^\infty
(-1)^k\e^{-(2S+1)\eta k}{\sinh m\eta k\over \sinh(2S+1)\eta k}\right),
\ee
in the latter limit the dressed energy is
\be
\epsilon(v)=4\sinh\eta\sum_{k=1}^\infty
\e^{-\eta k}{\sinh\eta k\over \sinh(2S+1)\eta k}(\cos k v-1).
\ee

\newpage
\def\and{and\ }
\def\eds{eds.\ }
\def\edi{ed.\ }

\references

\def\mtb{M. T. Batchelor}
\def\rjb{R. J. Baxter}
\def\dk{D. Kim}
\def\pap{P. A. Pearce}
\def\nyr{N. Yu. Reshetikhin}
\def\ak{A. Kl\"umper}

\refis{AbramS64} M. Abramowitz, I. A. Stegun, ``Handbook of 
Mathematical Functions", Washington, U.S. National Bureau of Standards 1964;
New York, Dover 1965.

\refis{Affl86} I. Affleck, \prl 56, 746, 1986

\refis{AKLT87} I. Affleck, T. Kennedy, E. H. Lieb \and H. Tasaki, \prl 59, 
799, 1987

\refis{AKLT88} I. Affleck, T. Kennedy, E. H. Lieb \and H. Tasaki, \cmp 115, 
477, 1988

\refis{AfflGSZ89} I. Affleck, D. Gepner, H. J. Schulz \and T. Ziman,
\jpa 22, 511, 1989

\refis{AkutDW89} Y. Akutsu, T. Deguchi \and M. Wadati, in Braid Group, Knot
Theory and Statistical
Mechanics, \eds C. N. Yang \and M. L. Ge, World Scientific, Singapore, 1989

\refis{AkutKW86a} Y. Akutsu, A. Kuniba \and M. Wadati,\jpj 55, 1466, 1986

\refis{AkutKW86b} Y. Akutsu, A. Kuniba \and M. Wadati,\jpj 55, 2907, 1986

\refis{AlcaBB87} F. C. Alcaraz, M. N. Barber \and \mtb,\prl 58, 771, 1987

\refis{AlcaBB88} F. C. Alcaraz, M. N. Barber \and \mtb,\annp 182, 280, 1988

\refis{AlcaBGR88} F. C. Alcaraz, M. Baake, U. Grimm \and V. Rittenberg,
\jpa 21, L117, 1988

\refis{AlcaM89}  F. C. Alcaraz \and M. J. Martins, \jpa 22, 1829, 1989

\refis{AlcaM90}  F. C. Alcaraz \and M. J. Martins, \jpa 23, 1439-51, 1990

\refis{Alex75} S. Alexander, \pla 54, 353-4, 1975

\refis{AndrBF84} G. E. Andrews, \rjb\ \and P. J. Forrester, \jsp 35, 193,
1984

\refis{And87} P. W. Anderson, Science 235, 1196, 1987

\refis{And90} P. W. Anderson, \prl 64, 1839, 1990

\refis{BDV82} O. Babelon, H. J. de Vega, \and C. M. Viallet, \npb 200 [FS4], 266, 1982

\refis{BarbBP87} \mtb, M.N. Barber \and \pap,\jsp 49, 1117, 1987

\refis{BarbB89} M.N. Barber \and \mtb, \prb 40, 4621, 1989

\refis{Barb91} M.N. Barber, \physica A 170, 221, 1991

\refis{BaresB90} P. A. Bares \and G. Blatter, \prl 64, 2567, 1990

\refis{Bariev8182} R. Z. Bariev, \tmp 49, 261, 1981; 1021, 1982

\refis{Bariev82} R. Z. Bariev, \tmp 49, 1021, 1982

\refis{Bariev91} R. Z. Bariev, \jpa 24, L549, 1991; L919, 1991

\refis{BarievKSZ93} R. Z. Bariev, A. Kl\"{u}mper, A. Schadschneider 
\and J. Zittartz, \jpa 26, 1249, 1993; 4863

\refis{Bariev94a} R. Z. Bariev, \prb 49, 1474, 1994

\refis{Bariev94b} R. Z. Bariev, submitted to J. Phys. A

\refis{BarouchM71} E. Barouch \and B. M. McCoy, \pra 3, 786, 1971

\refis{Baxt70} \rjb,\jmp 11, 3116, 1970

\refis{Baxt71b} \rjb,\prl 26, 834, 1971

\refis{Baxt72} \rjb,\annp 70, 193, 1972

\refis{Baxt73} \rjb,\jsp 8, 25, 1973

\refis{Baxt80} \rjb,\jpa 13, L61--70, 1980

\refis{Baxt81a} \rjb,\physica 106A, 18--27, 1981

\refis{Baxt81b} \rjb,\jsp 26, 427--52, 1981

\refis{Baxt82a} \rjb,\jsp 28, 1, 1982

\refis{Baxt82b} \rjb, ``Exactly Solved Models in Statistical Mechanics",
Academic Press, London, 1982.

\refis{BaxtP82} \rjb\space \and \pap,\jpa 15, 897, 1982

\refis{BaxtP83} \rjb\space \and \pap,\jpa 16, 2239, 1983

\refis{BazhR89} V.V. Bazhanov \and \nyr,\ijmpa 4, 115--42, 1989 

\refis{BazhB93} V.V. Bazhanov \and \rjb,\physica A 194, 390--396, 1993 

\refis{BednorzM86} J. G. Bednorz \and K. A. M"uller, \zpb 64, 189, 1986

\refis{BelaPZ84} A. A. Belavin, A. M. Polyakov \and A. B. Zamolodchikov,
\npb 241, 333, 1984

\refis{Bethe31} H. A. Bethe,\zp 71, 205, 1931

\refis{BlotCN86} H. W. J. Bl\"ote, J. L. Cardy \and M. P. Nightingale, \prl
56, 742,
1986

\refis{BogK89} N. M. Bogoliubov \and V. E. Korepin, \ijmpb 3, 427-439, 1989

\refis{BogIR86} N. M. Bogoliubov, A.\ G.\ Izergin \and 
N.\ Y.\ Reshetikhin, \jetpl 44, 405, 1986

\refis{BretzD71} M. Bretz \and J. G. Dash, \prl 27, 647, 1971

\refis{Bretz77} M. Bretz, \prl 38, 501, 1977

\refis{Buy86} W. J. L. Buyers, R. M. Morra, R. L. Armstrong, P. Gerlach
\and K. Hirakawa, \prl 56, 371, 1986

\refis{KawUO89} N. Kawakami, T. Usuki \and A. Okiji, \pla 137, 287, 1989

\refis{KawY90} N. Kawakami \and S.-K. Yang, \prl 65, 2309, 1990

\refis{KawY91} N. Kawakami \and S.-K. Yang, \prb 44, 7844, 1991

\refis{Kaw93} N. Kawakami, \prb 47, 2928, 1993

\refis{KorBI93} V. E. Korepin, N.M. Bogoliubov, \and A.G. Izergin,
``Quantum Inverse Scattering Method and Correlation
Functions", Cambridge University Press, 1993.

\refis{Morra88} R. M. Morra, W. J. L. Buyers, R. L. Armstrong \and K. Hirakawa,
\prb 38, 543, 1988

\refis{ShasS90} B. S. Shastry \and B. Sutherland, \prl 66, 243, 1990

\refis{Shastry88} B. S. Shastry, \jsp 50, 57, 1988

\refis{Stei87} M. Steiner, K. Kakurai, J. K. Kjems, D. Petitgrand \and R. Pynn,
\jappp 61, 3953, 1987

\refis{Tun90} Z. Tun, W. J. L. Buyers, R. L. Armstrong, K. Hirakawa \and
B. Briat, \prb 42, 4677, 1990

\refis{Tun91} Z. Tun, W. J. L. Buyers, A. Harrison \and J. A. Rayne, \prb 43,
13331, 1991

\refis{Ren87} J. P. Renard, M. Verdaguer, L. P. Regnault, W. A. C. Erkelens,
J. Rossa-Mignod \and W. G. Stirling, \eurolett 3, 945, 1987

\refis{Ren88} J. P. Renard, M. Verdaguer, L. P. Regnault, W. A. C. Erkelens,
J. Rossa-Mignod, J. Ribas, W. G. Stirling \and C. Vettier, \jappp 63, 3538, 1988

\refis{Reg89} L. P. Regnault, J. Rossa-Mignod, J. P. Renard, M. Verdaguer
\and C. Vettier, \physica B 156 \& 157, 247, 1989

\refis{Colom87} P. Colombet, S. Lee, G. Ouvrard \and R. Brec, \jcr, 134, 1987

\refis{deGroot82} H. J. M. de Groot, L. J . de Jongh, R. D. Willet \and
J. Reeyk, \jappp 53, 8038, 1982

\refis{Capp88} A. Cappelli, Recent Results in Two-Dimensional Conformal
Field
Theory, in Proceedings of the XXIV International Conference on High Energy
Physics,
\eds R. Kotthaus \and J. K\"uhn, Springer, Berlin, 1988

\refis{CappIZ87a} A. Cappelli, C. Itzykson \and J.-B. Zuber, \npb {280
[FS18]},
445--65, 1987

\refis{CappIZ87b} A. Cappelli, C. Itzykson \and J.-B. Zuber, \cmp 113,
1--26, 1987

\refis{Card84a} J. L. Cardy, \jpa 17, L385, 1984

\refis{Card86a} J. L. Cardy, \npb {270 [FS16]}, 186, 1986

\refis{Card86b} J. L. Cardy, \npb {275 [FS17]}, 200, 1986

\refis{Card88} J. L. Cardy, ``Phase Transitions and Critical Phenomena,
Vol.11",
\eds C. Domb \and J.L. Lebowitz, Academic Press, London 1988

\refis{Card89} J. L. Cardy, Conformal Invariance and Statistical Mechanics,
in Les
Houches, Session XLIV, Fields, Strings and Critical Phenomena, \eds E.
Br\'ezin \and
J. Zinn-Justin, 1989

\refis{ChoiKK90} J.-Y. Choi, K. Kwon \and D. Kim, \eurolett xx, to appear,
1990

\refis{ChoiKP89} J.-Y. Choi, D. Kim \and \pap, \jpa 22, 1661--71, 1989

\refis{CvetDS80} D. M. Cvetkovic, M. Doob \and H. Sachs, ``Spectra of
Graphs", Academic Press, London 1980

\refis{DateJKMO87} E. Date, M. Jimbo, A. Kuniba, T. Miwa, \and M. Okado,
\npb
B290, 231--273, 1987

\refis{DateJKMO88} E. Date, M. Jimbo, A. Kuniba, T. Miwa, \and M. Okado,
\aspm 16,
17, 1988

\refis{DateJMO86} E. Date, M. Jimbo, T. Miwa \and M. Okado,\lmp 12, 209,
1986

\refis{DateJMO87} E. Date, M. Jimbo, T. Miwa \and M. Okado,\prb 35, 2105--7,
1987

\refis{DaviP90} B. Davies \and \pap, \ijmpb {}, this issue, 1990

\refis{deVeK87} H. J. de Vega \and M. Karowski, \npb {285 [FS19]}, 619, 1987

\refis{deVeW85} H. J. de Vega \and F. Woynarovich,\npb 251, 439, 1985

\refis{deVeW90} H. J. de Vega \and F. Woynarovich,\jpa 23, 1613, 1990

\refis{DestdeVeW92} C. Destri \and H. J. de Vega,\prl 69, 2313, 1992

\refis{diFrSZ87} P. di Francesco, H. Saleur \and J.-B. Zuber, \jsp 49,
57--79, 1987

\refis{diFrZ89} P. di Francesco \and J.-B. Zuber, $SU(N)$ Lattice Models
Associated
with Graphs, Saclay preprint SPhT/89-92, 1989

\refis{DijkVV88} R. Dijkgraaf, E. Verlinde \and H. Verlinde, in Proceedings
of the
1987 Copenhagen Conference, World Scientific, 1988

\refis{DijkVVV89} R. Dijkgraaf, C. Vafa, E. Verlinde \and H. Verlinde,\cmp
123, 485, 1989

\refis{DombG76} ``Phase Transitions and Critical Phenomena, Vol.6",
Academic Press, London 1976

\refis{FateZ85} V. A. Fateev \and A. B. Zamolodchikov, \jetp 62, 215, 1985

\refis{FendG89} P. Fendley \and P. Ginsparg, \npb 324, 549--80, 1989

\refis{FodaN89} O. Foda \and B. Nienhuis, \npb {},{},1989

\refis{FoersK93} A. Foerster \and M. Karowski, \npb 408 [FS], 512, 1993

\refis{FrieQS84} D. Friedan, Z. Qiu \and S. Shenker, \prl 52, 1575, 1984; in
``Vertex Operators in Mathematics and Physics", \eds J. Lepowsky, S.
Mandelstam \and
I.M. Singer, Springer, 1984

\refis{FrahmK90} H. Frahm \and V. E. Korepin, \prb 42, 10553, 1990

\refis{FrahmYF90} H. Frahm, N.-C. Yu \and M. Fowler, \npb 336, 396, 1990

\refis{Frei93} W.-D. Freitag, Dissertation, Universit\"at zu K\"oln, 1993

\refis{GepnQ87} D. Gepner \and Z. Qiu, \npb 285, 423--53, 1987

\refis{Gins88} P. Ginsparg,\npb {295 [FS21]}, 153--70, 1988

\refis{Gins89a} P. Ginsparg, Applied Conformal Field Theory, in Les
Houches,
Session XLIV, Fields, Strings and Critical Phenomena, \eds E. Br\'ezin \and
J.
Zinn-Justin, 1989

\refis{Gins89b} P. Ginsparg, Some Statistical Mechanical Models and
Conformal Field
Theories, Trieste Spring School Lectures, HUTP-89/A027

\refis{GradR80} I.S. Gradshteyn \and I.M. Ryzhik, ``Tables of Integrals,
Series and Products", Academic
Press, New York, 1980.

\refis{Grif72} R. B. Griffiths, ``Phase Transitions and Critical Phenomena,
Vol.1",\eds C. Domb \and M. S. Green, Academic Press, London 1972

\refis{Hald83a} F. D. M. Haldane, \prl 50, 1153, 1983

\refis{Hald83b} F. D. M. Haldane, \pla 93, 464, 1983

\refis{Hame85} C. J. Hamer,\jpa 18, L1133, 1985

\refis{Hame86} C. J. Hamer,\jpa 19, 3335, 1986

\refis{Hirsch89a} J. E. Hirsch, \pla 134, 451, 1989

\refis{Hirsch89b} J. E. Hirsch, \physica  C 158, 326, 1989

\refis{Huse82} D. A. Huse,\prl 49, 1121--4, 1982

\refis{Huse84} D. A. Huse, \prb 30, 3908, 1984

\refis{Idz94} M. Idzumi, T. Tokihiro \and M. Arai, \jpI 4, 1151, 1994

\refis{ItzySZ88} C. Itzykson, H. Saleur \and J-B. Zuber, ``Conformal
Invariance and Applications to
Statistical Mechanics", World Scientific, Singapore, 1988

\refis{JimbM84} M. Jimbo \and T. Miwa, \aspm 4, 97--119, 1984

\refis{JimbMO87} M. Jimbo, T. Miwa \and M. Okado, \lmp 14, 123--31, 1987

\refis{JimbMO88} M. Jimbo, T. Miwa \and M. Okado, \cmp 116, 507--25, 1988

\refis{JimbMT89} M. Jimbo, T. Miwa \and A. Tsuchiya,``Integrable Systems in
Quantum Field Theory and
Statistical Mechanics", \aspm 19, ,1989

\refis{JohnKM73} J.D. Johnson, S. Krinsky, \and B.M. McCoy,\pra 8, 2526,
1973

\refis{Kac79} V. G. Kac, \lnp 94, 441--445, 1979

\refis{KadaB79} L. P. Kadanoff \and A. C. Brown, \annp 121, 318--42, 1979

\refis{Karo88} M. Karowski, \npb {300 [FS22]}, 473, 1988

\refis{Kato87} A. Kato, \mpla 2, 585, 1987

\refis{KimP87} \dk\space \and \pap,\jpa 20, L451--6, 1987

\refis{KimP89}  \dk\space \and \pap,\jpa 22, 1439--50, 1989

\refis{Kiri89} E. B. Kiritsis, \plb  217, 427, 1989

\refis{KiriR86} A. N. Kirillov \and N. Yu. Reshetikhin,\jpa 19, 565, 1986

\refis{KiriR87} A. N. Kirillov \and N. Yu. Reshetikhin,\jpa 20, 1565, 1987

\refis{KlassM90} T. R. Klassen \and E. Melzer, \npb 338, 485, 1990

\refis{KlassM91} T. R. Klassen \and E. Melzer, \npb 350, 635, 1991

\refis{KlumBiq} A. Kl\"{u}mper, \eurolett 9, 815, 1989; \jpa 23, 809, 1990

\refis{KlumB90} A. Kl\"{u}mper \and \mtb,\jpa 23, L189, 1990

\refis{KlumBP91} A. Kl\"{u}mper, \mtb \ \and \pap, \jpa 24, 3111--3133, 1991

\refis{KlumP91} A. Kl\"{u}mper \and \pap, \jsp 64, 13--76, 1991

\refis{Klum92c} A. Kl\"{u}mper, unver"offentlichte Rechnungen, (1992)

\refis{KlumZ88} A. Kl\"{u}mper \and J. Zittartz,\zpb 71, 495, 1988

\refis{KlumZ88App} A. Kl\"{u}mper \and J. Zittartz,\zpb 71, 495, 1988, 
Appendix A

\refis{KlumZ89} A. Kl\"{u}mper \and J. Zittartz,\zpb 75, 371, 1989

\refis{KlumZ8VM} A. Kl\"{u}mper \and J. Zittartz,\zpb 71, 495, 1988;
\zpb 75, 371, 1989

\refis{KlumSZ89} A. Kl\"{u}mper, A. Schadschneider \and J. Zittartz,
\zpb 76, 247, 1989

\refis{KlumSZMPG} A. Kl\"{u}mper, A. Schadschneider \and J. Zittartz,
\jpa 24, L955-L959, 1991; \zpb 87, 281-287, 1992

\refis{Klum89} \ak, \eurolett 9, 815, 1989

\refis{KlumP92} \ak\  \and \pap, \physica 183A, 304-350, 1992

\refis{Klum92} \ak , \Annp 1, 540, 1992

\refis{Klum93} \ak , \zpb 91, 507, 1993

\refis{Klum92b} \ak, in preparation

\refis{KlumWZ93} \ak, T. Wehner \and J. Zittartz, \jpa 26, 2815, 1993

\refis{Klum94} \ak , in Vorbereitung

\refis{KlumWeh94} \ak\ \and T. Wehner, in Vorbereitung

\refis{Knabe88} S. Knabe, \jsp 52, 627, 1988

\refis{Koma} T. Koma, \ptp 78, 1213, 1987; \bf 81, \rm 783, (1989)

\refis{KorepinS90} V. E. Korepin \and N. A. Slavnov, \npb 340, 759, 1990

\refis{ItsIK92} A. R. Its, A. G. Izergin \and V. E. Korepin, \physica D 54, 351, 1992

\refis{ItsIKS93} A. R. Its, A. G. Izergin, V. E. Korepin \and N. A. Slavnov, 
\prl 70, 1704, 1993

\refis{IKR89} A. G. Izergin, V. E. Korepin \and N. Yu. Reshetikhin, \jpa 22, 2615, 1989

\refis{KuniY88} A. Kuniba \and T. Yajima,\jsp 52, 829, 1988

\refis{KuliRS81} P. P. Kulish, N. Yu. Reshetikhin \and E. K. Sklyanin, \lmp
5, 393, 1981 

\refis{Kuniba92} A. Kuniba, ``Thermodynamics of the $U_q(X_r^{(1)})$ Bethe Ansatz System with $q$ a Root of Unity", ANU preprint (1991)    

\refis{LeeS88} K. Lee \and P. Schlottmann, \jpcoll 49 C8, 709, 1988

\refis{Lewi58} L. Lewin, Dilogarithms and Associated Functions, MacDonald,
London, 1958

\refis{LiebWu68} E. H. Lieb \and F. Y. Wu, \prl 20, 1445, 1968

\refis{LiebWu72} E. H. Lieb \and F. Y. Wu, 
``Phase Transitions and Critical Phenomena,
Vol.1",
\eds C. Domb \and M. S. Green, Academic Press, London 1988

\refis{LutherP74} A. Luther \and I. Peschel, \prb 9, 2911, 1974

\refis{Martins91} M. J. Martins, \prl 22, 419, 1991 and private communication 
(1991)

\refis{Muell} E. M\"uller-Hartmann, unpublished results, (1989)

\refis{Muell89} E. M\"uller-Hartmann, unver"offentlichte Ergebnisse, (1989)

\refis{Mura89} J. Murakami, \aspm 19, 399--415, 1989

\refis{Nien87} B. Nienhuis, in Phase Transitions and Critical Phenomena,
Vol.11,
\eds C. Domb \and J.L. Lebowitz, Academic Press, 1987

\refis{NahmRT92} W. Nahm, A. Recknagel \and M. Terhoven, Preprint ``Dilogarithm
identities in conformal field theory'', 1992

\refis{NighB86} M. P. Nightingale \and H. W. J. Bl"ote, \prb 33, 659, 1986

\refis{OwczB87} A. L. Owczarek \and \rjb,\jsp 49, 1093, 1987

\refis{ParkBiq} J. B. Parkinson,\jpc 20, L1029, 1987; \jpc 21, 3793, 1988; 
\jphc 8, 1413, 1988

\refis{ParkB85} J. B. Parkinson \and J. C. Bonner, \prb 32, 4703, 1985

\refis{PaczP90} I. D. Paczek \and J. B. Parkinson,\jpcon 2, 5373, 1990

\refis{Pasq87a} V. Pasquier,\npb {285 [FS19]}, 162, 1987

\refis{Pasq87b} V. Pasquier,\jpa 20, {L217, L221}, 1987

\refis{Pasq87c} V. Pasquier,\jpa 20, {L1229, 5707}, 1987

\refis{Pasq88} V. Pasquier,\npb {B295 [FS21]}, 491--510, 1988

\refis{Pear85} \pap,\jpa 18, 3217--26, 1985

\refis{Pear87prl} \pap,\prl 58, 1502--4, 1987

\refis{Pear87jpa} \pap,\jpa 20, 6463--9, 1987

\refis{Pear90ijmpb} \pap,\ijmpb 4, 715--34, 1990

\refis{PearB90} \pap\space \and \mtb, \jsp 60, 77--135, 1990

\refis{PearK87} \pap \and \dk, \jpa, 20, 6471-85, 1987

\refis{PearS88} \pap\space \and K. A. Seaton,\prl 60, 1347, 1988

\refis{PearS89} \pap\space \and K. A. Seaton,\annp 193, 326, 1989

\refis{PearS90} \pap\space \and K. A. Seaton,\jpa 23, 1191--1206, 1990

\refis{Pear91} \pap, Row Transfer Matrix Functional Equations for
$A$--$D$--$E$ Lattice Models,
 to be published, 1991

\refis{PearK91} \pap\space \and A. Kl\"umper, \prl 66, 974, 1991

\refis{Pear92} \pap, \ijmpa 7, Suppl.1B, 791, 1992

\refis{PerkS81} J. H. H. Perk \and C. L. Schultz, \pla 84, 407, 1981

\refis{Resh83jetp} \nyr,\jetp 57, 691, 1983

\refis{Resh83lmp} \nyr, \lmp 7, 205--13, 1983

\refis{Sale88} H. Saleur, Lattice Models and Conformal Field Theories, in
Carg\`ese
School on Common Trends in Condensed Matter and Particle Physics, 1988

\refis{SaleB89} H. Saleur \and M. Bauer, \npb 320, 591--624, 1989

\refis{SaleD87} H. Saleur \and P. di Francesco, Two Dimensional Critical
Models on a
Torus, in Brasov Summer School on Conformal Invariance and String Theory,
1987

\refis{Samuel73} E. J. Samuelson, \prl 31, 936, 1973

\refis{Schlott87} P. Schlottmann, \prb 36, 5177, 1987

\refis{Schlott92} P. Schlottmann, \jpc 4, 7565, 1992

\refis{Schul83} C. L. Schultz, \physica 122A, 71, 1983

\refis{SeatP89} K. A. Seaton \and \pap\space, \jpa 22, 2567--76, 1989

\refis{Strog79} Yu. G. Stroganov, \pla 74, 116, 1979

\refis{Suth70} B. Sutherland, \jmp 11, 3183, 1970

\refis{Suth75} B. Sutherland, \prb 12, 3795, 1975

\refis{Suzuki85} M. Suzuki, \prb 31, 2957, 1985

\refis{SuzukiI87} M. Suzuki \and M. Inoue, \ptp 78, 787, 1987

\refis{Suzuki87} M. Suzuki, in ``Quantum Monte Carlo Methods in 
Equilibrium and Nonequilibrium Systems",
\edi M. Suzuki, Springer Verlag, 1987

\refis{SuzukiAW90} J. Suzuki, Y. Akutsu \and M. Wadati, \jpj 59, 2667-2680, 1990

\refis{SuzukiNW92} J. Suzuki, T. Nagao \and M. Wadati, \ijmpb 6, 1119, 1992

\refis{Tak71} M. Takahashi, \ptp 46, 401, 1971

\refis{TakTBA} M. Takahashi, \ptp 46, 401, 1971; \ptp 50, 1519, 1973

\refis{Tak91} M. Takahashi, \prb 43, 5788, 1991; \prb 44, 12382, 1991

\refis{Tak91a} M. Takahashi, \prb 43, 5788, 1991

\refis{Tak91b} M. Takahashi, \prb 44, 12382, 1991

\refis{TempL71} H. N. V. Temperley \and E. H. Lieb, \prs 322, 251, 1971

\refis{Tetel82} M. G. Tetel'man, \jetp 55, 306, 1982

\refis{TruS83} T. T. Truong \and K. D. Schotte, \npb 220, 77, 1983

\refis{Tsun91} H. Tsunetsugu, \jpj 60, 1460, 1991

\refis{vonGR87} G. von Gehlen \and V. Rittenberg, \jpa 20, 227, 1987

\refis{WadaDA89} M. Wadati, T. Deguchi \and Y. Akutsu, \prep 180, 247--332,
1989

\refis{Woyn87} F. Woynarovich, \prl 59, 259, 1987

\refis{WoynE87} F. Woynarovich \and H.-P. Eckle, \jpa 20, L97, 1987

\refis{Yang69} C. N. Yang \and C. P. Yang, \jmp 10, 1115, 1969

\refis{Yang66} C. N. Yang \and C. P. Yang, \pr 147, 303, 1966; 150, 321

\refis{Yang62} C. N. Yang, \rmp 34, 691, 1962

\refis{Yang67} C. N. Yang, \prl 19, 1312, 1967

\refis{CPYang67} C. P. Yang, \prl 19, 586, 1967

\refis{YangG89} C. N. Yang \and M. L. Ge (Editors), Braid Group, Knot Theory
and Statistical
Mechanics, World Scientific, Singapore, 1989

\refis{ZamoF80} A. B. Zamolodchikov \and V. Fateev, \sjnp 32, 198, 1980

\refis{Zamo80} A. B. Zamolodchikov, \jetp 52, 325, 1980; \cmp 79, 489, 1981

\refis{Zamo91} Al. B. Zamolodchikov, \plb 253, 391--4, 1991; \npb 358, 
497--523, 1991

\refis{Zamo91a} Al. B. Zamolodchikov, \plb 253, 391--4, 1991

\refis{Zamo91b} Al. B. Zamolodchikov, \npb 358, 497--523, 1991

\refis{ZhangR88} F. C. Zhang and T. M. Rice, \prb 37, 3759, 1988

\endreferences

\vskip0.5cm

\section*{Captions}

\begin{itemize}

\item[Figure 1:] (a) Depiction of the critical exponent $\alpha$ for $S=1/2$ 
and different values of $\eta=$ 0.1, 0.5, 1, 2, 10. 
For values of $\alpha$ above
the broken line we have dominating superconducting correlations. 
(b) The same for $S=1$.

\item[Figure 2:] (a) The charge stiffness $D_c$ for $S=1/2$ and
different anisotropies $\eta=$ 0.1, 0.5, 1, 2, 5. (b) The same for
$S=1$. Note that the values for $\eta=5$ are zero within the 
graphical resolution.

\item[Figure 3:] (a) The effective mass $m$ of 
the charge carriers for $S=1/2$ and
different anisotropies $\eta=$ 0.1, 0.5, 1, 2. 
(b) The same for $S=1$.
\end{itemize}

\newpage

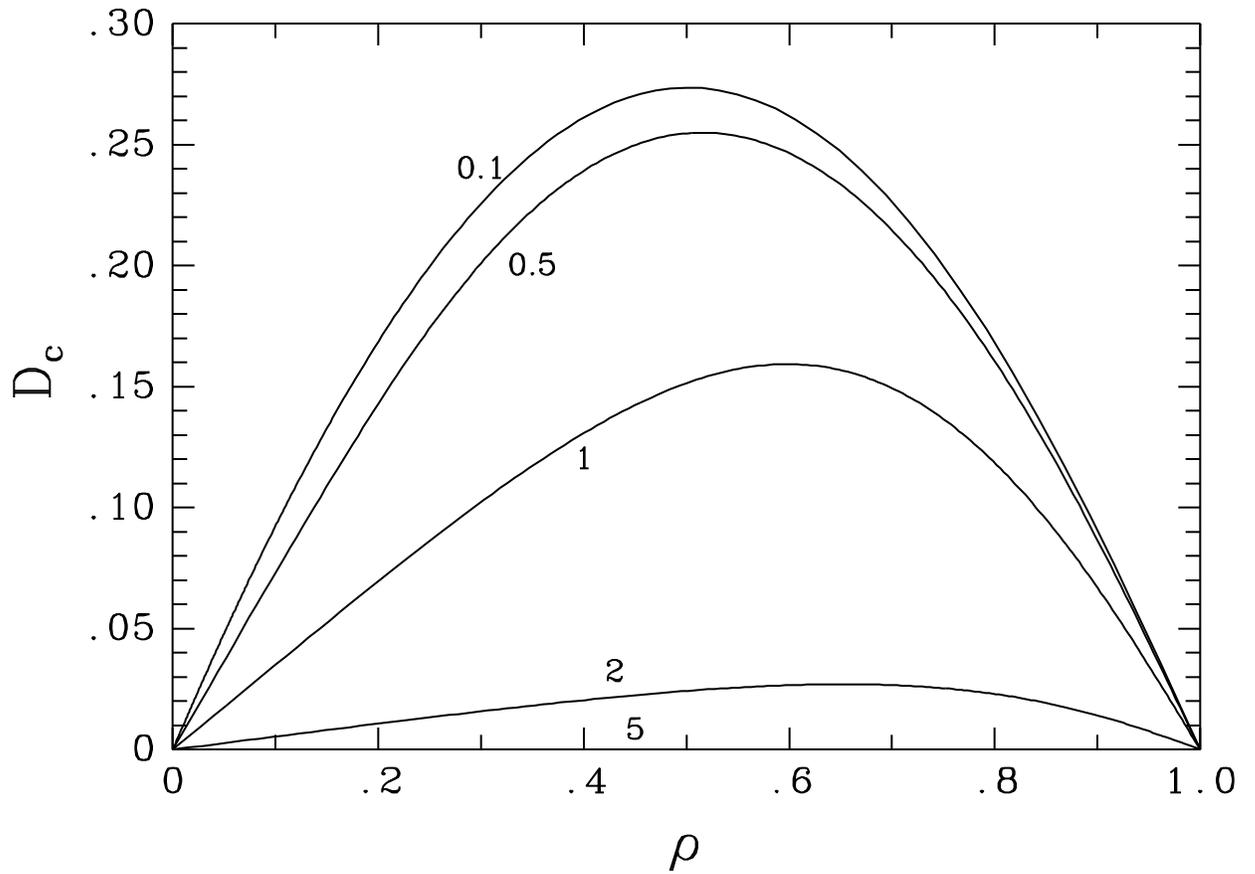
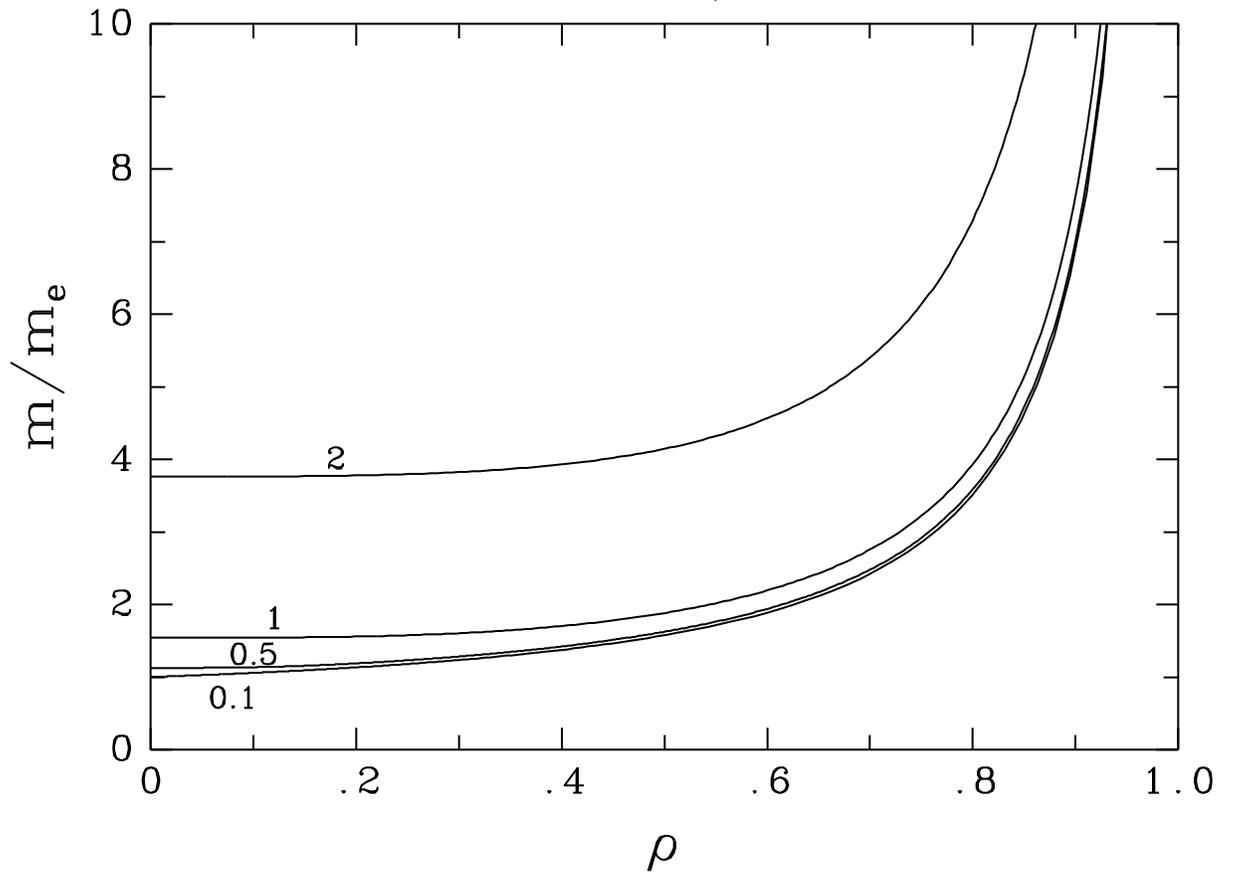
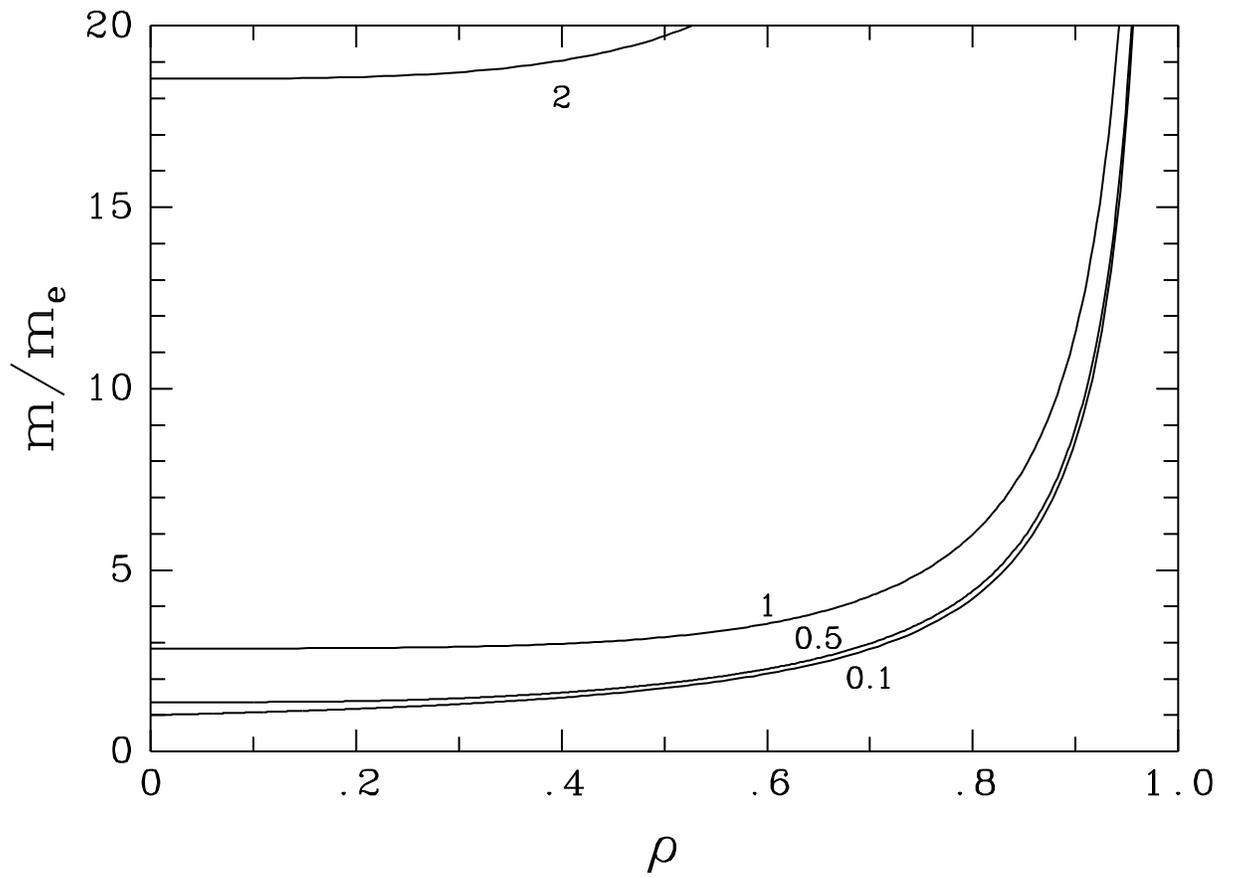
\end{document}

***************************************************************************

 It follows the uuencoded and compressed tar-file for the postcript
 files of the 3 figures.
 (do: uudecode figures.uu, uncompress figures.tar.Z, tar -xvf figures.tar)

***************************************************************************